**Eleven shear surface waves guided by inclusions in magneto-electro-elastic materials**

Arman Melkumyan [a]

*Department of Mechanics, Yerevan State University, Alex Manoogyan Str. 1, Yerevan 375025, Armenia*

Eleven pure shear surface waves are shown to exist in magneto-electro-elastic materials, which contain different plane inclusions inside them. The plane inclusions affect the electric and magnetic fields only, so that the medium remains mechanically continuous across it. Expressions for velocities of propagation of these 11 surface waves are obtained in explicit forms. It is expected that these waves will have several applications in surface acoustic wave devices.

[a] Electronic mail: melk_arman@yahoo.com

Bleustein[1] and Gulyaev[2] have shown that an elastic shear surface wave can be guided by the free surface of piezoelectric materials in class 6 mm, and then Danicki[3] has described the propagation of a shear surface wave guided by an embedded conducting plane. In this paper we consider an infinite magneto-electro-elastic[4-8] medium which contains embedded infinitesimally thin plane with different electric and magnetic properties and study the possibilities for surface waves to be guided by it. The medium can also be considered as a system of two perfectly bonded magneto-electro-elastic half-spaces with infinitesimally thin layers in the plane of bonding. Discussing different magneto-electrical boundary conditions in the plane of bonding we obtain surface waves with eleven different velocities.

Let $x_1$, $x_2$, $x_3$ denote rectangular Cartesian coordinates with $x_3$ oriented in the direction of the sixfold axis of a magneto-electro-elastic material in class 6 mm. Introducing electric potential $\varphi$ and magnetic potential $\phi$, so that $E_1 = -\varphi_{,1}$, $E_2 = -\varphi_{,2}$, $H_1 = -\phi_{,1}$, $H_2 = -\phi_{,2}$, the five partial differential equations which govern the mechanical displacements $u_1$, $u_2$, $u_3$, and the potentials $\varphi$, $\phi$, reduce to two sets of equations when motions independent of the $x_3$ coordinate are considered. The equations of interest in the present paper are those governing the $u_3$ component of the displacement and the potentials $\varphi$, $\phi$, and can be written in the form

$$c_{44}\nabla^2 u_3 + e_{15}\nabla^2 \varphi + q_{15}\nabla^2 \phi = \rho \ddot{u}_3 ,$$
$$e_{15}\nabla^2 u_3 - \varepsilon_{11}\nabla^2 \varphi - d_{11}\nabla^2 \phi = 0 , \qquad (1)$$
$$q_{15}\nabla^2 u_3 - d_{11}\nabla^2 \varphi - \mu_{11}\nabla^2 \phi = 0 ,$$

where $\nabla^2$ is the two-dimensional Laplacian operator, $\nabla^2 = \partial^2/\partial x_1^2 + \partial^2/\partial x_2^2$, $\rho$ is the mass density, $c_{44}$, $e_{15}$, $\varepsilon_{11}$, $q_{15}$, $d_{11}$ and $\mu_{11}$ are elastic, piezoelectric, dielectric, piezomagnetic, electromagnetic and magnetic constants, and the superposed dot indicates differentiation with respect to time. The constitutive equations which relate the stresses $T_{ij}$ ($i, j = 1, 2, 3$), the electric displacements $D_i$ ($i = 1, 2, 3$) and the magnetic induction $B_i$ ($i = 1, 2, 3$) to $u_3$, $\varphi$ and $\phi$ are

$T_1 = T_2 = T_3 = T_{12} = 0$, $D_3 = 0$, $B_3 = 0$ ,

$T_{23} = c_{44}u_{3,2} + e_{15}\varphi_{,2} + q_{15}\phi_{,2}$ , $T_{13} = c_{44}u_{3,1} + e_{15}\varphi_{,1} + q_{15}\phi_{,1}$ ,

$D_1 = e_{15}u_{3,1} - \varepsilon_{11}\varphi_{,1} - d_{11}\phi_{,1}$ , $D_2 = e_{15}u_{3,2} - \varepsilon_{11}\varphi_{,2} - d_{11}\phi_{,2}$ , (2)

$B_1 = q_{15}u_{3,1} - d_{11}\varphi_{,1} - \mu_{11}\phi_{,1}$ , $B_2 = q_{15}u_{3,2} - d_{11}\varphi_{,2} - \mu_{11}\phi_{,2}$ .

Solving Eqs. (1) for $\nabla^2 u_3$, $\nabla^2 \varphi$ and $\nabla^2 \phi$ we find that after defining functions $\psi$ and $\chi$ by

$$\psi = \varphi - \frac{e_{15}\mu_{11} - q_{15}d_{11}}{\varepsilon_{11}\mu_{11} - d_{11}^2} u_3 , \quad \chi = \phi - \frac{q_{15}\varepsilon_{11} - e_{15}d_{11}}{\varepsilon_{11}\mu_{11} - d_{11}^2} u_3 , \tag{3}$$

the solution of Eqs. (1) is reduced to the solution of

$$\nabla^2 u_3 = \rho \tilde{c}_{44}^{-1} \ddot{u}_3 , \quad \nabla^2 \psi = 0 , \quad \nabla^2 \chi = 0 , \tag{4}$$

where

$$\tilde{c}_{44} = c_{44} + \left(e_{15}^2 \mu_{11} - 2e_{15}q_{15}d_{11} + q_{15}^2 \varepsilon_{11}\right)/\left(\varepsilon_{11}\mu_{11} - d_{11}^2\right)$$

$$= \overline{c}_{44}^e + \varepsilon_{11}^{-1} \left(d_{11}e_{15} - q_{15}\varepsilon_{11}\right)^2/\left(\varepsilon_{11}\mu_{11} - d_{11}^2\right)$$

$$= \overline{c}_{44}^m + \mu_{11}^{-1} \left(d_{11}q_{15} - e_{15}\mu_{11}\right)^2/\left(\varepsilon_{11}\mu_{11} - d_{11}^2\right) \tag{5}$$

is magneto-electro-elastically stiffened elastic constant, $\overline{c}_{44}^e = c_{44} + e_{15}^2/\varepsilon_{11}$ is piezoelectrically stiffened elastic constant and $\overline{c}_{44}^m = c_{44} + q_{15}^2/\mu_{11}$ is piezomagnetically stiffened elastic constant. With the analogy to the piezoelectric coupling factor $k_e^2 = e_{15}^2/(\varepsilon_{11}\overline{c}_{44}^e)$ and the piezomagnetic coupling factor $k_m^2 = q_{15}^2/(\mu_{11}\overline{c}_{44}^m)$ introduce piezoelectromagnetic coupling factor

$$k_{em}^2 = \tilde{c}_{44}^{-1}\left(e_{15}^2\mu_{11} - 2e_{15}q_{15}d_{11} + q_{15}^2\varepsilon_{11}\right)/\left(\varepsilon_{11}\mu_{11} - d_{11}^2\right)$$

$$= e_{15}^2/(\varepsilon_{11}\tilde{c}_{44}) + \tilde{c}_{44}^{-1}\varepsilon_{11}^{-1}\left(d_{11}e_{15} - q_{15}\varepsilon_{11}\right)^2/\left(\varepsilon_{11}\mu_{11} - d_{11}^2\right)$$

$$= q_{15}^2/(\mu_{11}\tilde{c}_{44}) + \tilde{c}_{44}^{-1}\mu_{11}^{-1}\left(d_{11}q_{15} - e_{15}\mu_{11}\right)^2/\left(\varepsilon_{11}\mu_{11} - d_{11}^2\right). \tag{6}$$

Using the introduced functions $\psi$ and $\chi$ and the magneto-electro-elastically stiffened elastic constant, the constitutive Eqs. (2) can be written in the following form:

$$T_{23} = \tilde{c}_{44} u_{3,2} + e_{15}\psi_{,2} + q_{15}\chi_{,2} , \quad T_{13} = \tilde{c}_{44} u_{3,1} + e_{15}\psi_{,1} + q_{15}\chi_{,1} ,$$

$$D_1 = -\varepsilon_{11}\psi_{,1} - d_{11}\chi_{,1} , \quad D_2 = -\varepsilon_{11}\psi_{,2} - d_{11}\chi_{,2} , \tag{7}$$

$$B_1 = -d_{11}\psi_{,1} - \mu_{11}\chi_{,1} , \quad B_2 = -d_{11}\psi_{,2} - \mu_{11}\chi_{,2} .$$

Introduce short notations $e = e_{15}$, $\mu = \mu_{11}$, $d = d_{11}$, $\varepsilon = \varepsilon_{11}$, $q = q_{15}$, $c = c_{44}$, $\overline{c}^e = \overline{c}_{44}^e$, $\overline{c}^m = \overline{c}_{44}^m$, $\tilde{c} = \tilde{c}_{44}$, $w = u_3$, $T = T_{23}$, $D = D_2$, $B = B_2$ and use subscripts $A$ and $B$ to refer to the half-spaces $x_2 > 0$ and $x_2 < 0$, respectively. As the materials in the half spaces $x_2 > 0$ and $x_2 < 0$ are identical, one has that $e_A = e_B = e$,

$\mu_A = \mu_B = \mu$, $d_A = d_B = d$, $\varepsilon_A = \varepsilon_B = \varepsilon$, $q_A = q_B = q$, $c_A = c_B = c$, $\overline{c}_A^e = \overline{c}_B^e = \overline{c}^e$, $\overline{c}_A^m = \overline{c}_B^m = \overline{c}^m$, $\tilde{c}_A = \tilde{c}_B = \tilde{c}$.

The conditions at infinity require that

$w_A$, $\varphi_A$, $\phi_A \to 0$ as $x_2 \to \infty$,

$w_B$, $\varphi_B$, $\phi_B \to 0$ as $x_2 \to -\infty$, (8)

and the mechanical bonding conditions require that

$w_A = w_B$, $T_A = T_B$ on $x_2 = 0$. (9)

Consider the possibility of a solution of Eqs. (3)-(4) of the form

$w_A = w_{0A} \exp(-\xi_2 x_2) \exp[i(\xi_1 x_1 - \omega t)]$,

$\psi_A = \psi_{0A} \exp(-\xi_1 x_2) \exp[i(\xi_1 x_1 - \omega t)]$, (10)

$\chi_A = \chi_{0A} \exp(-\xi_1 x_2) \exp[i(\xi_1 x_1 - \omega t)]$,

in the half space $x_2 > 0$ and of the form

$w_B = w_{0B} \exp(\xi_2 x_2) \exp[i(\xi_1 x_1 - \omega t)]$,

$\psi_B = \psi_{0B} \exp(\xi_1 x_2) \exp[i(\xi_1 x_1 - \omega t)]$, (11)

$\chi_B = \chi_{0B} \exp(\xi_1 x_2) \exp[i(\xi_1 x_1 - \omega t)]$,

in the half space $x_2 < 0$. These expressions satisfy the conditions (8) if $\xi_1 > 0$ and $\xi_2 > 0$; the second and the third of Eqs. (4) are identically satisfied and the first of Eqs. (4) requires

$\tilde{c}(\xi_1^2 - \xi_2^2) = \rho \omega^2$. (12)

Now the elastic contact conditions (9) together with different magneto-electrical contact conditions on $x_2 = 0$ must be satisfied. The following cases of magneto-electrical contact conditions on $x_2 = 0$ are of our interest in the present paper:

1a) $\varphi_A = \varphi_B$, $D_A = D_B$, $\phi_A = \phi_B = 0$;

1b) $D_A = D_B = 0$, $\phi_A = \phi_B = 0$;

2a) $\varphi_A = \varphi_B = 0$, $\phi_A = \phi_B$, $B_A = B_B$;

2b) $\varphi_A = \varphi_B = 0$, $B_A = B_B = 0$;

3) $\varphi_A = \varphi_B = 0$, $\phi_A = \phi_B = 0$;

4) $D_A = D_B = 0$, $B_A = 0$, $\phi_B = 0$;

5) $\varphi_A = 0$, $D_B = 0$, $B_A = B_B = 0$;

6) $D_A = 0$, $\varphi_B = 0$, $B_A = 0$, $\phi_B = 0$;

7) $\varphi_A = 0$, $D_B = 0$, $\phi_A = \phi_B = 0$; (13)

8) $\varphi_A = \varphi_B = 0$, $B_A = 0$, $\phi_B = 0$;

9) $\varphi_A = 0$, $D_B = 0$, $B_A = 0$, $\phi_B = 0$;

10) $\varphi_A = \varphi_B$, $D_A = D_B$, $B_A = 0$, $\phi_B = 0$;

11) $\varphi_A = 0$, $D_B = 0$, $\phi_A = \phi_B$, $B_A = B_B$;

12a) $\varphi_A = \varphi_B$, $D_A = D_B$, $\phi_A = \phi_B$, $B_A = B_B$;

12b) $D_A = D_B = 0$, $\phi_A = \phi_B$, $B_A = B_B$;

12c) $\varphi_A = \varphi_B$, $D_A = D_B$, $B_A = B_B = 0$;

12d) $D_A = D_B = 0$, $B_A = B_B = 0$.

Each of the 17 groups of conditions in Eqs. (13) together with Eqs. (9), Eqs. (10)-(11) leads to a system of six homogeneous algebraic equations for $w_{0A}$, $\psi_{0A}$, $\chi_{0A}$, $w_{0B}$, $\psi_{0B}$, $\chi_{0B}$, the existence of nonzero solution of which requires that the determinant of that system must be equal to zero. This condition for the determinant together with Eq. (12) determines the surface wave velocities $V_s = \omega/\xi_1$. In the case of 1a) of Eqs. (13) this procedure leads to a surface wave with velocity

$$V_{s1}^2 = (\tilde{c}/\rho)\left(1 - \left[k_{em}^2 - e^2/(\tilde{c}\varepsilon)\right]^2\right). \tag{14}$$

The same velocity is obtained in the case 1b). The cases 2a) and 2b) both lead to a surface wave with velocity

$$V_{s2}^2 = (\tilde{c}/\rho)\left(1 - \left[k_{em}^2 - q^2/(\tilde{c}\mu)\right]^2\right). \tag{15}$$

Each of the cases 3) to 11) leads to its own one surface wave and the corresponding surface wave velocities are

$$V_{s3}^2 = (\tilde{c}/\rho)\left(1 - k_{em}^4\right);$$

$$V_{s4}^2 = (\tilde{c}/\rho)\left(1 - \tfrac{1}{4}\left[k_{em}^2 - e^2/(\tilde{c}\varepsilon)\right]^2\right);$$

$$V_{s5}^2 = (\tilde{c}/\rho)\left(1 - \tfrac{1}{4}\left[k_{em}^2 - q^2/(\tilde{c}\mu)\right]^2\right);$$

$$V_{s6}^2 = (\tilde{c}/\rho)\left(1 - \tfrac{1}{4}k_{em}^4\right);$$

$$V_{s7}^2 = (\tilde{c}/\rho)\left(1 - \left[k_{em}^2 - \tfrac{1}{2}e^2/(\tilde{c}\varepsilon)\right]^2\right); \quad (16)$$

$$V_{s8}^2 = (\tilde{c}/\rho)\left(1 - \left[k_{em}^2 - \tfrac{1}{2}q^2/(\tilde{c}\mu)\right]^2\right);$$

$$V_{s9}^2 = (\tilde{c}/\rho)\left(1 - \left[k_{em}^2 - \frac{1}{2}\left(\frac{e^2}{\tilde{c}\varepsilon} + \frac{q^2}{\tilde{c}\mu}\right)\right]^2\right);$$

$$V_{s10}^2 = (\tilde{c}/\rho)\left(1 - \left(\frac{\varepsilon\mu}{2\varepsilon\mu - d^2}\right)^2\left[k_{em}^2 - e^2/(\tilde{c}\varepsilon)\right]^2\right);$$

$$V_{s11}^2 = (\tilde{c}/\rho)\left(1 - \left(\frac{\varepsilon\mu}{2\varepsilon\mu - d^2}\right)^2\left[k_{em}^2 - q^2/(\tilde{c}\mu)\right]^2\right).$$

The cases 12a) to 12d) do not lead to any surface wave.

If the magneto-electro-elastic material degenerates to a piezoelectric material, so that $q \to 0$, $d \to 0$, the surface waves that have velocities $V_{s1}$, $V_{s4}$, $V_{s10}$ disappear and

$$V_{s2}, V_{s3}, V_{s8} \to V_{bg} = \sqrt{(\bar{c}^e/\rho)(1 - k_e^4)};$$

$$V_{s5}, V_{s6}, V_{s7}, V_{s9}, V_{s11} \to \sqrt{(\bar{c}^e/\rho)(1 - \tfrac{1}{4}k_e^4)}, \quad (17)$$

where $V_{bg}$ is the Bleustein-Gulyaev surface wave speed, so that the number of surface waves decreases from 11 to 2 when the magneto-electro-elastic material is changed to a piezoelectric material.

[1] J.L. Bleustein, Appl. Phys. Lett. **13**, 412 (1968).

[2] Y.V. Gulyaev, JETP Lett. **9**, 37 (1968).

[3] E. Danicki, Appl. Phys. Lett. **64**, 969 (1994).

[4] C.W. Nan, Phys. Rev. B **50**, 6082 (1994).

[5] K. Srinivas, G. Prasad, T. Bhimasankaram and S. V. Suryanarayana, Modern Phys. Lett. B **14**, 663 (2000).

[6] K. Mori and M. Wuttig, Appl. Phys. Lett. **81**, 100 (2002).

[7] P. Yang, K. Zhao, Y. Yin, J.G. Wan, and J.S. Zhu, Appl. Phys. Lett. **88**, 172903 (2006).

[8] S. Srinivas, J.Y. Li, Y.C. Zhou, A.K. Soh, J. Appl. Phys. **99**, 043905 (2006).